\documentclass[fleqn]{article}

\usepackage{amssymb}
\usepackage{amsmath}
\usepackage[dvips]{graphicx}

\begin{document}

\title{A method to calculate Franck-Condon factors in terms of the tomographic probability representation}
\author{E. D. Zhebrak \\
Moscow Institute of Physics and Technology, Institutsky lane 9,\\
Dolgoprudny, Russia}
\maketitle

\begin{abstract}
We introduce a new method to calculate Franck-Condon factors in polyatomic molecules that is based on the tomographic probability approach to quantum mechanics. This approach is implemented to calculate transition probabilities in various systems under an instantaneous change of frequency and equilibrium position of nuclei in a molecule by an external force. The problem is considered for different types of the Dushinsky matrix and for any quantity of atoms in a molecule.
\end{abstract}

\section{Introduction}
Vibrational spectra study in polyatomic molecules is now the problem of great interest because of the numerous applications in different fundamental and applied branches: in optical properties of crystals investigation, in planetary atmospheres spectroscopy etc. The recent progress of experimental techniques that allows to make high-resolution measurements and operate with compound polyatomic molecules, for example with DNA and its bases (\cite{1}-\cite{4}) formed a need for effective methods of vibrational spectra analysis.\\

In the molecule radiation and absorbtion study transition probabilities between the initial and the final states are usually determined by the Franck-Condon (FC) factors that are defined as an absolute square of an initial and a final wave function overlap integral. This approach is suitable in the Born-Oppenheimer approximation and requires that nuclear coordinates change very little during the electronic transitions and also supposes that the vibrations in the molecule are of small amplitude and consequently that the system states can be represented in the scope of harmonic oscillator model. Despite of simplicity of the considered approach it requires efficient calculation schemes being applied to polyatomic molecules.\\

The first to introduce the FC integral calculation approach for multidimensional systems were Sharp and Rosenstock \cite{5} who derived an elegant relation for the overlap integrals using the generating function method. Several years later Doktorov et al. (\cite{6}-\cite{7}) proposed a new method based on the coherent states approach that also gave widely implemented recurring expressions for the FC factors. Further the identity of the two mentioned approaches was proved in \cite{8}. A recent study of various molecular vibronic transitions in terms of coherent state-based generating function method was observed in detail in \cite{8a}.\\

In the FC integrals the integrated expression in the wave function representation has the following form: a factor depending on the initial and the final vibrational level numbers, an exponential part and a $3N'-6$ - dimensional Hermite polynomial or $3N'-5$ - dimensional one in the case of a linear molecule (where $N'$ is the number of atoms). The main problem in FC integrals numerical calculation is that the integrated expression in the general case can't be factorized because of the Dushinsky effect \cite{9} appearing in the dependence of the electronic final normal coordinates on the initial ones. Furthermore, for the full study of vibrational spectrum at a certain energy $E=n\omega $ one needs to compute $C_{n}^{3N'-6+n-1} $ FC integrals. It means that by the FC integrals calculation for polyatomic molecules one has to operate with the integrals of a huge dimension and by the wide enough energy range also has to store in memory a significant number of them. Moreover the integrated expression is not fully positive.\\

Because of the mentioned problems a number of new FC integrals calculation methods are having been introduced. Among the recent papers one should mention the following ones \cite{10}-\cite{18}. The methods proposed in \cite{10} and \cite{11} are based on the relevant vibronic transition selection and taking into account the corresponding FC factors only. In the works \cite{12} and \cite{13} the problem of extreme dimension of FC integrals was solved by factorizing the integrated expression. It was achieved by representing the Dushinsky rotation matrix (in \cite{12}) or the matrix depending on the coefficients before the squared normal coordinates in the initial and the final wave functions (in \cite{13}) in a block-diagonal form. In \cite{14} vibronic spectra of polycyclic aromatic hydrocarbon molecules was investigated by obtaining Franck-Condon factors within the damped harmonic oscillator method. Instead of direct calculation of the overlap integral the intensities of absorption and fluorescence lines were found with the help of the Huang-Rhys parameters. They were get by scaling unperturbed and diagonalizing perturbed Hessian matrices using empirically introduced scaling parameters.  An original approach was introduced in \cite{15} where the problem was solved in curvilinear normal coordinates.\\

There are also several recent works focused on the FC integral properties by the low dimensions. In \cite{16} the explicit formula for the FC factor in a two-dimensional system was obtained using the properties of Hermite polynomials and Gaussian states. In \cite{17},\cite{18} a new expression for the FC factor containing partial derivatives of the exponential part instead of the Hermite polynomials was obtained in terms of the contour integral method. The explicit forms of this expression were represented for the two- and three- dimensional cases.\\

More detailed information about the recently introduced approaches to FC integrals calculation, their advantages and implementation one can find, for example, in the following review: \cite{19}.\\

Nevertheless, the wave function approach is not the only method to determine quantum states which can be also represented in terms of various phase-space functions. Among the phase-space approaches the most useful one for molecule vibration transitions study is the Wigner function representation \cite{Wigner}. The Wigner treatment of triatomic and triatomic-like photodissociation dynamics was developed by S. Goursaud et al. (\cite{Goursaud}), R. Brown and E. Heller (\cite{BH}), M. Sheppard and R. Walker (\cite{SW}) and others. The recent work \cite {Gonzalez} gives an excellent review of the Wigner function application in molecular spectra analysis and also proposes new semiclassical Wigner representation based methods that allow to observe not only vibrational but also rotational distributions.  \\
Wigner function has a form of a phase space distribution corresponding to a certain quantum state that makes it close to classical mechanics. But in fact it can't be a real probability distribution because Wigner function depends simultaneously on coordinate and momentum that correspond to non-commuting operators. It leads to the fact that Wigner distribution can have negative values. On the other side there is a representation mapping a quantum state on a function that has a sense of probability distribution. This approach called tomographic interpretation of quantum mechanics was introduced in \cite{23} and has been well developed during the last years. In terms of this approach, quantum states are defined by the specific function called tomogram, that depends on a coordinate in a rotated and squeezed phase space and on parameters of the rotation and squeezing. It enables not to include non-commuting operators that leads to the fact that the tomogram is real, fully positive and has all other properties of probability distribution. By this reason, such kind of problems like transition probability calculation in different physical systems can be solved in terms of the tomographic approach in the most essential way. However the applicability of the tomographic formalism (that is well-known in quantum optics) to molecule vibrational spectra analysis has never been discussed.\\

This paper is aimed to give a new method of vibronic spectra study based on the FC factors calculation in terms of the tomographic probability approach. This method is shown to be also applicable for vibrational dynamics investigation in polyatomic molecules because it requires integration only of a real and positive function that can be crucial for computing FC factors of a high dimension.\\

The article is organized as follows: in the Section 2 is given a brief insight into the problem of transition probabilities calculation in terms of wave function, Wigner function and the tomographic probability approach. The Section 3 contains a general description of the proposed method and discussion of its application to some simple particular cases. In the Section 4 we apply the mentioned approach to the problem of Franck-Condon factors computation in an arbitrary polyatomic molecule.

\section{Molecule vibrational transition probabilities}

\subsection{Transition probability calculation in terms of wave functions}

In a molecule vibrational transition from the initial state $\left| n  \right\rangle$ to the final state $\left| m  \right\rangle$ leads to radiation (or can be caused by absorbtion) at the frequency $\omega=\hbar^{-1}(E_m-E_n)$. The intensity of the radiation (absorbtion) depends on the matrix element of a dipole moment $\left\langle n \right|\mu\left|m\right\rangle$ that in the adiabatic approximation can be expressed as
\begin{equation}
\left\langle n \right|\mu\left|m\right\rangle = \left\langle n\right|R_{nm}(\vec{q})\left|m\right\rangle \approx \left\langle n\right|R_{nm}(\vec{q_0})\left|m\right\rangle = R_{nm}(\vec{q_0})\left\langle n\right|m\rangle.
\end {equation}
The line's intensity is thus proportional to the modulus squared of the wave functions overlap integral called Franck-Condon factor:
\begin{equation} \label{e2}
I_{nm}\sim\left|\left\langle n\right|m\rangle\right|^2.
\end {equation}
This form coincides with the definition of transition probability in a quantum system stated by the Born rule.

\subsection{Transition probabilities in terms of Wigner functions}

During the development of the quantum mechanics appeared different approaches to define quantum states. Primarily there were functions on the phase space including quasiprobability distribution called the Wigner function \cite{Wigner} that is the Fourier transform of density matrix in the position representation:
\begin{equation} \label{eq6}
W\left(q,p\right)= \int \exp \left(-i p\xi \right){\left\langle q+\frac{\xi}{2}  \right|} \rho {\left| q-\frac{\xi}{2}  \right\rangle}  d\xi .
\end{equation}
Here and below we assume that $\hbar=1$. Transition probability between two quantum states of a system can be also found in terms of the Wigner function. It equals the overlap integral of the Wigner functions defining the initial state $W_{n} \left(q,p\right)$ and the final state $W_{m} \left(q,p\right)$ as it was shown in \cite{22}:
\begin{equation} \label{eq7}
P_{nm} =\frac{1}{2\pi}\int W_{n}  \left(q,p\right)W_{m} \left(q,p\right)dqdp.
\end{equation}

This relation was proved in \cite{SW} to define Franck-Condon factors in molecules by spontaneous vibronic transitions including photodissociation processes. But as is well known the Wigner function can have negative values and by this reason it cannot be interpreted as probability distribution.

\subsection{Transition probabilities in terms of tomographic representation}
The function on the phase space caring the whole information about a quantum state and having the meaning of probability distribution was first introduced in the work \cite{23}. This function $w\left(X,\mu ,\nu \right)$ called the symplectic tomogram depends on three variables: $X=\mu q+\nu p$ is the coordinate in a squeezed and rotated phase space and  $\mu $, $\nu $ are the real parameters of this squeeze and rotation. The particular case of the symplectic tomogram appearing by $\mu=cos{\theta}$ and $\nu=sin{\theta}$ is called the optical tomogram. This representation is rather common in optics because the optical tomogram is directly measurable in homodyne detection experiments (see, for example \cite{24}, \cite{25}) that enables to avoid a complicated procedure of quantum state reconstruction. This is one more reason to try to extend tomographic probability approach to experimental domains that deal with quantum state reconstruction. Due to these advantages tomographic approach has been well developed in application to a number of problems including the problem of transition probability calculation (see, for example \cite{26}, \cite{27}). A sufficiently complete review of the tomographic approach, its meaning and applicability was made by A. Ibort et. al. in \cite{Ibort}.\\
The symplectic tomogram is related to the Wigner function by the reversible Radon transform
\begin{equation} \label{eq8}
w\left(X,\mu ,\nu \right)=\frac{1}{2\pi } \int W\left(q,p\right)\delta \left(X-\mu q-\nu p\right)dqdp
\end{equation}
and because of \eqref{eq6} has a one-valued relation to the wave function:
\begin{equation} \label{eq9}
w\left(X,\mu ,\nu \right)=\frac{1}{2\pi \left|\nu \right|} \left|\int \psi \left(y\right)e^{\frac{i\mu }{2\nu } y^{2} -\frac{iX}{\nu } y} dy \right|^{2}.
\end{equation}
From the definition \eqref{eq8} also follow the main properties of the symplectic tomogram, namely its nonnegativity:
\begin{equation}\nonumber
w\left(X,\mu ,\nu \right)\ge{0},
\end{equation}
integrability:
\begin{equation}\nonumber
\int w\left(X,\mu ,\nu \right)dX<\infty
\end{equation}
and homogeneity:
\begin{equation}\nonumber
w\left(\lambda X,\lambda \mu ,\lambda \nu \right)=\frac{1}{\left|\lambda\right|}w\left(X,\mu ,\nu \right).
\end{equation}
 As it was shown in \cite{28} from the expressions \eqref{eq7} and \eqref{eq8} follows the relation for transition probability in terms of symplectic tomograms:
\begin{equation} \label{eq10}
P_{nm} =\frac{1}{2\pi } \int w_n \left(X,\mu ,\nu \right) w_m \left(Y,-\mu ,-\nu \right)e^{i\left(X+Y\right)} dXdYd\mu d\nu .
\end{equation}
Since (\ref{eq7}) defines Franck-Condon factors in a molecule corresponding to transition from the eigenstate $\left|n,0\right\rangle$ to the eigenstate $\left|m,t\right\rangle$ after a perturbation, the relation (\ref{eq10}) can be apparently used by the same problem. This method of transition probability calculation will be applied in the following paragraphs to the problems of transitions between the energy levels of nuclei in a polyatomic molecule.

\section {Franck-Condon factors in polyatomic molecules}

Inside the Born-Oppenheimer approximation when the dipole moment alternation can be considered to be small one can use the both relations \eqref{e2} and \eqref{eq10} to calculate Franck-Condon factors. Since in this case we can consider the problem as transitions between the energy levels of a harmonic oscillator, the Franck-Condon factors in polyatomic molecules can be obtained as transition probabilities in multidimensional oscillators under perturbation.\\

In a polyatomic molecule according to the Dushinsky effect the normal coordinates of the electrons after the perturbation are represented as a linear combination of the normal coordinates before it: $\vec{q'}=\Lambda\vec{q}+\vec{\gamma}$.\\
Before considering the general case we will give a short observation of particular cases when the Dushinsky rotation matrix $\Lambda$ is an identity matrix (a shift of an equilibrium position of nuclei in a molecule) and when the translation vector $\vec{\gamma}$ equals zero (a change of frequency).

\subsection {Particular case: shift of an equilibrium position of nuclei in a molecule}

In this chapter we will assume that the molecule wave functions before and after a shift of an equilibrium position have the following forms in the harmonic oscillator approximation:
\begin {equation}
\psi \left(x,t_0\right)=\frac{1}{\sqrt{2^n n!}\sqrt[{4}]{\pi}} e^{-\frac{x^2}{2}}H_n\left(x\right)
\end {equation}
and
\begin {equation}
\psi \left(x,t_{fin}\right)=\frac{1}{\sqrt{2^m m!}\sqrt[{4}]{\pi}} e^{-\frac{\left(x+\gamma\right)^2}{2}}H_m\left(x+\gamma\right)
\end {equation}
In terms of wave functions the problem of an instantaneous shift of an equilibrium position in harmonic oscillator was first discussed by J. Schwinger in \cite{29}. In the relation obtained in the scope of this work for the one-dimensional case the transition probabilities were expressed in a specific form with the help of Laguerre polynomials:
\begin{equation} \label{eq3_1}
P_{nm} =\frac{n_{<} !}{n_{>} !} \exp \left(-\left|\kappa \right|^{2} \right)\left|\kappa \right|^{2\left|m-n\right|} \left(L_{n_{<} } ^{\left|m-n\right|} \left(\left|\kappa \right|^{2} \right)\right)^{2} ,
\end{equation}
where $n_{<} =\min \left(n,m\right)$, $n_{>} =\max \left(n,m\right)$, $\kappa=\frac{1}{\sqrt{2\omega}}\int_{t_0}^{t_{fin}} f\left(t\right)e^{-i\omega t}dt$, $\omega$ is the vibrational frequency and $f\left( t\right)$ is the external force.\\

The same problem in terms of the tomographic probability approach was solved in \cite{27}.
Having obtained the symplectic tomographies according to \eqref{eq9} one can get Franck-Condon factors in a diatomic molecule under a shift of an equilibrium position of its nuclei with the help of \eqref{eq10} in the following form:
\begin{eqnarray} \label{eq3_3}
\nonumber\ {P_{nm} =\frac{1}{2\pi } \int \frac{1}{2^{n} n!2^{m} m!\pi \left(\nu ^{2} +\mu ^{2} \right)} e^{-\frac{X^{2} }{\nu ^{2} +\mu ^{2} } -\frac{\left(Y-\gamma\mu \right)^{2} }{\nu ^{2} +\mu ^{2} } } e^{i\left(X+Y\right)}} \\
\times \left|H_{n} \left(\frac{X}{\sqrt{\nu ^{2} +\mu ^{2} } } \right)H_{m} \left(\frac{Y-\gamma \mu }{\sqrt{\nu ^{2} +\mu ^{2} } } \right)\right|^{2}  dXdYd\mu d\nu .
\end{eqnarray}
Comparison with the Schwinger's result \eqref{eq3_1} gives a new relation for the Franck-Condon factors expressed in terms of the shift value $\gamma $:
\begin{equation} \label{eq3_4}
P_{nm} =\frac{n_{<} !}{n_{>} !} \left(\frac{\gamma ^{2} }{2} \right)^{\left|m-n\right|} \exp \left(-\frac{\gamma ^{2} }{2} \right)\left(L_{n_{<} } ^{\left|m-n\right|} \left(\frac{\gamma ^{2} }{2} \right)\right)^{2} .
\end{equation}
The relation \eqref{eq3_3} can be easily generalized for the case of a $N'$-atomic molecule:
\begin{eqnarray}
\nonumber P_{n_{1} ...n_{N} m_{1} ...m_{N} } =\frac{1}{4^{N} \pi ^{N} 2^{n_{1} +...+n_{N} +m_{1} +...+m_{N} } n_{1} !...n_{N} !m_{1} !...m_{N} !}\\
\nonumber\int \frac{1}{\left|\nu _1 ...\nu _N  \right|^2}\frac{1}{\left|\det \left(\frac{E}{2}-\Omega\right)\right|^2}\exp \left[i\left(\vec{X}-\vec{Y}\right)\right]\\
\nonumber\times\exp\left[{\frac{1}{4}\vec{\eta}^{T}\left(\frac{E}{2}-\Omega\right)^{-1}}\vec{\eta}+
\frac{1}{4}\left(\vec{\tilde{\eta}}-\vec{\gamma}\right)^{T}\left(\frac{E}{2}-\Omega\right)^{-1}\left(\vec{\tilde{\eta}}-\vec{\gamma}\right)
-\frac{1}{2}\vec{\gamma}^{T}\vec{\gamma}\right]\\
\times \left|H_{_{\vec{n}} }^{\vartheta} \left(\frac{1}{2}\Omega^{-1}\vec{\eta}\right)H_{_{\vec{m}} }^{\vartheta}\left(\vartheta^{-1} \left(\vec{\gamma}+\frac{1}{2} \left(\frac{E}{2}-\Omega\right) ^{-1} \left(\vec{\tilde{\eta}}-\vec{\gamma}\right)\right)\right)\right|^{2} d\vec{X}d\vec{Y}d\vec{\mu }d\vec{\nu },
\end{eqnarray}
where $N=3N'-6$ (or $N=3N'-5$ for linear molecules), $\vartheta=E-\left(E-2\Omega\right)^{-1}$, $E$ is an identity matrix, $\vec{\eta }$ is a $N$ - dimensional vector determined by the tomogram parameters
$\vec{\eta }=\left(-\frac{iX_{1} }{\nu _{1} } ,-\frac{iX_{2} }{\nu _{2} } ,...,-\frac{iX_{N} }{\nu _{N} } \right)$ and $\Omega $ is a $N\times N$ diagonal matrix with $\left(\Omega _{kk} \right)=\frac{i\mu _{k} }{2\nu _{k} } $, $\left(\Omega _{jk} \right)_{j\ne k} =0$.\\

\subsection {Particular case: an instantaneous change of frequency of nuclei in a molecule }

In this chapter we will consider the case of an instantaneous shift of the nuclei oscillation frequency. Apparently the symplectic tomograms of the initial and the final states have the same form under the different oscillator lengths $l=\sqrt{\frac{\hbar }{m\omega } } $ and $l'=\sqrt{\frac{\hbar }{m\omega '} } $:
\begin{equation} \label{eq3_5}
w_{n}^{in} \left(X,\mu ,\nu ,l\right)=\frac{l}{2^{n} n!\sqrt{\pi \left(\nu ^{2} +\mu ^{2} l^{4} \right)} } e^{-\frac{X^{2} l^{2} }{\nu ^{2} +\mu ^{2} l^{4} } } \left|H_{n} \left(\frac{Xl}{\sqrt{\nu ^{2} +\mu ^{2} l^{4} } } \right)\right|^{2}
\end{equation}
and
\begin{equation} \label{eq3_6}
w_{n}^{fin} \left(X,\mu ,\nu ,l'\right)=\frac{l'}{2^{n} n!\sqrt{\pi \left(\nu ^{2} +\mu ^{2} l'^{4} \right)} } e^{-\frac{X^{2} l'^{2} }{\nu ^{2} +\mu ^{2} l'^{4} } } \left|H_{n} \left(\frac{Xl'}{\sqrt{\nu ^{2} +\mu ^{2} l'^{4} } } \right)\right|^{2} .
\end{equation}
The Franck-Condon factor corresponding to transition between the states $n$ and $m$ meets the relation
\begin{eqnarray} \label{eq3_7}
\nonumber {P_{nm} =\frac{ll'}{2^{n+m+1} \pi ^{2} } \int \frac{1}{\sqrt{\left(\nu ^{2} +\mu ^{2} l^{4} \right)\left(\nu ^{2} +\mu ^{2} l'^{4} \right)}}} \\
\nonumber \times \exp \left[-X^{2} \left(\frac{l^{2} }{\nu ^{2} +\mu ^{2} l^{4} } \right)+iX-Y^{2} \left(\frac{l'^{2} }{\nu ^{2} +\mu ^{2} l'^{4} } \right)+iY\right] \\
\times H_{n} ^{2} \left(\frac{Xl}{\sqrt{\nu ^{2} +\mu ^{2} l^{4} } } \right)H_{m} ^{2} \left(\frac{Xl'}{\sqrt{\nu ^{2} +\mu ^{2} l'^{4} } } \right)dXdYd\mu d\nu.
\end{eqnarray}
The calculation of the modulus squared of the wave functions overlap integral enables to express the transition probability in terms of 2-dimensional Hermite polynomials:
\begin{equation} \label{eq3_8}
P_{nm} =\frac{1}{2^{n+m} n!m!ll'} \left(\frac{2l^{2} l'^{2} }{l^{2} +l'^{2} } \right)\left|\frac{l^{2} -l'^{2} }{l^{2} +l'^{2} } \right|^{n+m} \left(H_{nm}^{\left\{\zeta \right\}} \left(0,0\right)\right)^{2} ,
\end{equation}
where $\zeta =\frac{2ll'}{l^{2} +l'^{2} } $.\\

Using the fact that the 2-dimensional Hermite polynomials are connected with the Legendre polynomials by the following relation \cite{31}:
\begin{equation} \label{eq3_8.1}
H_{nm}^{\left\{\zeta \right\}} \left(0,0\right)=n_{<} !\left(\zeta ^{2} -1\right)^{\frac{m+n}{4} } \left(-1\right)^{n_{>} +\frac{\left|n-m\right|}{4} } P_{m\max }^{m\min } \left(\frac{\zeta }{\sqrt{\zeta ^{2} -1} } \right)\\,
\end{equation}
where $n_{<} =\min \left(n,m\right)$,$n_{>} =\max \left(n,m\right)$, $m\min =\frac{\left|m-n\right|}{2} $, $m\max =\frac{m+n}{2}, $\\

one can obtain the Franck-Condon factor expressed with the help of the Legendre polynomials:
\begin{equation} \label{eq3_9}
P_{nm} =\frac{\left(n_{<} !\right)^{2} }{2^{n+m} n!m!} \zeta \left(P_{m\max }^{m\min } \left(\zeta \right)\right)^{2}.
\end{equation}
As is easy to see the parameter $\zeta $ can be also defined as the ratio of the geometric and the arithmetic means of the initial and the final oscillation frequencies of nuclei in a molecule:
\begin{equation} \label{eq3_10}
\zeta=\frac{\sqrt{\omega \omega '} }{\left(\frac{\omega +\omega '}{2} \right)} .
\end{equation}
When $n=m$ the relation \eqref{eq3_9} can be reduced to the following form:
\begin{equation} \label{eq3_11}
P_{nn} =\frac{1}{4^{n} } \zeta \left(P_{n} \left(\zeta \right)\right)^{2} .
\end{equation}
For the case of a $N'$ - atomic molecule one gets the following expression:
\begin{eqnarray}
\nonumber\ {P_{n_{1} ...n_{N} m_{1} ...m_{N} } =\frac{1}{4^{N} \pi ^{N} 2^{n_{1} +...+n_{N} +m_{1} +...+m_{N} } n_{1} !...n_{N} !m_{1} !...m_{N} !}} \\
\nonumber\int \frac{1}{\left|\nu _{1} ...\nu _{N} \right|^{2} \left|\det \left(\frac{\Lambda^2}{2}-\Omega\right)\right|^2}\exp \left[i\left(\vec{X}-\vec{Y}\right)\right]  \\
\nonumber\times \left|\exp \left[\frac{1}{4}\vec{\eta }^{T} \left(\frac{E}{2} -\Omega \right)^{-1}\vec{\eta}+\frac{1}{4}\vec{\tilde{\eta}}^{T}\left(\frac{\Lambda^2}{2}-\Omega\right)^{-1}\vec{\tilde{\eta}}\right]\right|^{2} \\
\nonumber \times \left|H_{_{\vec{n}} }^{\left\{E-\left(E-2\Omega\right)\right\}^{-1}} \left(\left(E-\left(E-2\Omega\right)^{-1}\right)^{-1} \left(E-2\Omega\right)^{-1}\vec{\eta}\right)\right|^{2} \\
\times \left|H_{_{\vec{m}} }^{\left\{E-\left(E-2\Omega\right)^{-1}\right\}} \left(\left(E-\left(E-2\Omega\right)^{-1}\right)^{-1} \left(\frac{1}{2}\Lambda\left(\frac{E}{2}-\Omega\right)^{-1}\vec{\eta}\right)\right)\right|^{2} d\vec{X}d\vec{Y}d\vec{\mu }d\vec{\nu }.
\end{eqnarray}

\section {Franck-Condon factors in an arbitrary polyatomic molecule}

\noindent Now we start considering an arbitrary $N'$-atomic molecule that can be represented as a $N$ - dimensional system with a quadratic Hamiltonian determined by the wave function
\begin{equation} \label{eq4_1}
\psi _{n_{1} ...n_{N} } \left(\vec{x}\right)=\frac{1}{\sqrt[{4}]{\pi ^{N} } \sqrt{2^{n_{1} +...+n_{N} } n_{1} !...n_{N} !} } \exp \left[-\frac{1}{2} \vec{x}^{T} \sigma \vec{x}+\vec{\varpi }^{T} \vec{x}+\varphi \right]H_ {\vec{n}}\left(\vec{x}\right)
\end{equation}
The phase $\varphi $ plays here the role of a normalizing factor.\\

Like in a one-dimensional case the wave function of this system after an instantaneous change of an equilibrium position and frequency of nuclei will have the same form as the wave function of the initial state with a scaled and shifted argument:
\begin{eqnarray}
\nonumber \tilde{\psi }_{n_{1} ...n_{N} } \left(\Lambda \vec{x}+\vec{\gamma}\right)=\frac{1}{\sqrt[{4}]{\pi ^{N} } \sqrt{2^{n_{1} +...+n_{N} } n_{1} !...n_{N} !} } \\
\times \exp \left[-\frac{1}{2} \left(\Lambda \vec{x}+\vec{\gamma}\right)^{T} \tilde{\sigma }\left(\Lambda \vec{x}+\vec{\gamma}\right)+\vec{\tilde{\varpi }}^{T} \left(\Lambda \vec{x}+\vec{\gamma}\right)+\tilde{\varphi }\right]H_ {\vec{n}}\left(\Lambda \vec{x}+\vec{\gamma}\right),
\end{eqnarray}
where the $N$ - dimensional vector $\vec{\gamma}$ and the $N\times N$ - matrix $\Lambda $ define the perturbation.\\

Using the direct calculation scheme \eqref{eq9} we obtain symplectic tomograms for the state after a perturbation:
\begin{eqnarray} \label{eq4_3}
\nonumber w_{n_{1} ...n_{N} } \left(\vec{X},\vec{\mu },\vec{\nu }\right)=\frac{1}{2^{N} \pi ^{\frac{N}{2} } 2^{n_{1} +...+n_{N} } n_{1} !...n_{N} !\left|\nu _{1} ...\nu _{N} \right|\left|\det \xi \right|}\\
\nonumber \times \left|\exp \left[\frac{1}{4} \vec{\kappa }^{T} \xi ^{-1} \vec{\kappa }-\frac{1}{2} \vec{\gamma}^{T} \tilde{\sigma }\vec{\gamma}+\vec{\tilde{\varpi }}^{T} \vec{\gamma}+\tilde{\varphi }\right]\right|^{2} \\
\times \left|H_{_{n_{1} ...n_{N} } }^{\left\{\tilde{\sigma }-\frac{1}{2} \tilde{\sigma }\Lambda \xi ^{-1} \Lambda \tilde{\sigma }\right\}} \left(\left(\tilde{\sigma }-\frac{1}{2} \tilde{\sigma }\Lambda \xi ^{-1} \Lambda \tilde{\sigma }\right)^{-1} \left(\tilde{\sigma }\vec{\gamma}+\frac{1}{2} \tilde{\sigma }\Lambda \xi ^{-1} \vec{\kappa }\right)\right)\right|^{2} ,
\end{eqnarray}
where $\xi =\frac{\Lambda \sigma \Lambda }{2} -\Omega $, $\vec{\kappa }=\Lambda \vec{\tilde{\varpi }}+\vec{\eta }-\Lambda \tilde{\sigma }\vec{\gamma}$,  $\vec{\eta }$ is a $N$ -dimensional vector determined by the tomogram parameters $\vec{\eta }=\left(-\frac{iX_{1} }{\nu _{1} } ,-\frac{iX_{2} }{\nu _{2} } ,...,-\frac{iX_{N} }{\nu _{N} } \right)$ and $\Omega $ is a $N\times N$ diagonal matrix with $\left(\Omega _{kk} \right)=\frac{i\mu _{k} }{2\nu _{k} } $, $\left(\Omega _{jk} \right)_{j\ne k} =0$.\\

The tomogram of the initial state has the same form under the following parameters:\\

$\tilde{\sigma }\to \sigma $, $\tilde{\varpi }\to \varpi $, $\vec{\gamma}=0$ and $\Lambda $ is an identity matrix.\\

According to \eqref{eq10} the Franck-Condon factor corresponding to transition between the state ${\left| n_{1} ...n_{N}  \right\rangle} $ before the perturbation and the state ${\left| m_{1} ...m_{N}  \right\rangle} $ after it is defined in terms of symplectic tomograms by the integral relation:
\begin{eqnarray} \label{eq4_4}
\nonumber\ {P_{n_{1} ...n_{N} m_{1} ...m_{N} } =\frac{1}{4^{N} \pi ^{N} 2^{n_{1} +...+n_{N} +m_{1} +...+m_{N} } n_{1} !...n_{N} !m_{1} !...m_{N} !}} \\
\nonumber\int \frac{1}{\left|\nu _{1} ...\nu _{N} \right|^{2} \left|\det \xi \det \tilde{\xi }\right|}\exp \left[i\left(\vec{X}-\vec{Y}\right)\right]  \\
\nonumber\times \left|\exp \left[\frac{1}{4} \left(\vec{\varpi }+\vec{\eta }\right)^{T} \left(\frac{\sigma }{2} -\Omega \right)^{-1} \left(\vec{\varpi }+\vec{\eta }\right)+\frac{1}{4} \vec{\tilde{\kappa }}^{T} \xi ^{-1} \vec{\tilde{\kappa }}-\frac{1}{2} \vec{\gamma}^{T} \tilde{\sigma }\vec{\gamma}+\vec{\tilde{\varpi }}^{T} \vec{\gamma}+\varphi +\tilde{\varphi }\right]\right|^{2} \\
\nonumber \times \left|H_{_{\vec{n}} }^{\left\{\sigma -\sigma \left(\sigma -2\Omega \right)^{-1} \sigma \right\}} \left(\left(\sigma -\sigma \left(\sigma -2\Omega \right)^{-1} \sigma \right)^{-1} \left(\sigma \left(\sigma -2\Omega \right)^{-1} \left(\vec{\varpi }+\vec{\eta }\right)\right)\right)\right|^{2} \\
\times \left|H_{_{\vec{m}} }^{\left\{\tilde{\sigma }-\frac{1}{2} \tilde{\sigma }\Lambda \xi ^{-1} \Lambda \tilde{\sigma }\right\}} \left(\left(\tilde{\sigma }-\frac{1}{2} \tilde{\sigma }\Lambda \xi ^{-1} \Lambda \tilde{\sigma }\right)^{-1} \left(\tilde{\sigma }\vec{\gamma}+\frac{1}{2} \tilde{\sigma }\Lambda \xi ^{-1} \vec{\tilde{\kappa }}\right)\right)\right|^{2} d\vec{X}d\vec{Y}d\vec{\mu }d\vec{\nu },
\end{eqnarray}
where $\vec{\tilde{\kappa }}=\Lambda \vec{\tilde{\varpi }}+\vec{\tilde{\eta }}-\Lambda \tilde{\sigma }\vec{\gamma}$ and $\vec{\tilde{\eta }}=\left(\frac{iY_{1} }{\nu _{1} } ,\frac{iY_{2} }{\nu _{2} } ,...,\frac{iY_{N} }{\nu _{N} } \right)$.\\

To obtain a formula for the transition probability determined above we apply the relation \eqref{e2} for wave functions taking into account the following integral property of multidimensional Hermite polynomials that were first derived in \cite{31}:
\begin{eqnarray} \label{eq4_5}
\nonumber \int H_{\vec{n}}^{\left\{S\right\}} \left(\vec{x}\right) H_{\vec{m}}^{\left\{T\right\}} \left(\Lambda \vec{x}+\vec{\gamma}\right)\exp \left[-\vec{x}^{T} M\vec{x}+\vec{c}^{T} \vec{x}\right]d\vec{x}\\
=\frac{\pi ^{\frac{N}{2} } }{\sqrt{\det M} } \exp \left(\frac{1}{4} cM^{-1} c\right)H_{nm}^{\left\{R\right\}} \left(\vec{y}\right),
\end{eqnarray}
where $R$ is a $\left(2N\times 2N\right)$ - block matrix determined as follows:
\[R=\left(\begin{array}{cc} {R_{11} } & {R_{12} } \\ {\tilde{R}_{12} } & {R_{22} } \end{array}\right),\]
\[R_{11} =S-\frac{1}{2} SM^{-1} S, R_{22} =T-\frac{1}{2} T\Lambda M^{-1} \Lambda ^{T} T, \tilde{R}_{12} =-\frac{1}{2} T\Lambda M^{-1} S\]
and $\vec{y}$ is a $2N$ - vector expressed in the following form:
\[\vec{y}=R^{-1} \left(\begin{array}{c} {\vec{z}_{1} } \\ {\vec{z}_{2} } \end{array}\right),\]
where $\vec{z}_{1} =\frac{1}{4} \left(SM^{-1} +M^{-1} S\right)\vec{c}$, $\vec{z}_{2} =\frac{1}{4} \left(T\Lambda M^{-1} +M^{-1} \Lambda ^{T} T\right)\vec{c}+T\vec{\gamma}$.\\

Using \eqref{eq4_5} it is easy to see that the Franck-Condon factor corresponding to the transition between the state ${\left| n_{1} ...n_{N}  \right\rangle} $ before the perturbation and the state ${\left| m_{1} ...m_{N}  \right\rangle} $ after it is expressed in terms of $2N$- dimensional Hermite polynomials:
\begin{eqnarray} \label{eq4_6}
\nonumber\ {P_{n_{1} ...n_{N} m_{1} ...m_{N} } =\frac{1}{2^{n_{1} +...+n_{N} +m_{1} +...+m_{N} -N} n_{1} !...n_{N} !m_{1} !...m_{N} !\left|\det \left(\sigma +\Lambda \tilde{\sigma }\Lambda \right)\right|}} \\ \times\left|H_{\vec{n}\vec{m}}^{\left\{R\right\}} \left(\vec{y}\right)e^{\frac{1}{2} \left(\vec{\varpi }+\Lambda \vec{\tilde{\varpi }}-\Lambda \tilde{\sigma }\vec{\gamma}\right)^{T} \left(\sigma +\Lambda \tilde{\sigma }\Lambda \right)^{-1} \left(\vec{\varpi }+\Lambda \vec{\tilde{\varpi }}-\Lambda \tilde{\sigma }\vec{\gamma}\right)-\frac{1}{2} \vec{\gamma}^{T} \tilde{\sigma }\vec{\gamma}+\vec{\tilde{\varpi }}^{T} \vec{\gamma}+\varphi +\tilde{\varphi }}\right|^{2} ,
\end{eqnarray}
where $R_{11} =\sigma -\frac{1}{2} \sigma \left(\sigma +\Lambda \tilde{\sigma }\Lambda \right)^{-1} \sigma $, $\tilde{R}_{12} =-\frac{1}{2} \tilde{\sigma }\Lambda \left(\sigma +\Lambda \tilde{\sigma }\Lambda \right)^{-1} \sigma $, $R_{22} =\tilde{\sigma }-\frac{1}{2} \tilde{\sigma }\Lambda \left(\sigma +\Lambda \tilde{\sigma }\Lambda \right)^{-1} \Lambda \tilde{\sigma }$, $\tilde{z}_{1} =\frac{1}{4} \left(\sigma \left(\sigma +\Lambda \tilde{\sigma }\Lambda \right)^{-1} +\left(\sigma +\Lambda \tilde{\sigma }\Lambda \right)^{-1} \sigma \right)\left(\vec{\varpi }+\Lambda \vec{\tilde{\varpi }}-\Lambda \tilde{\sigma }\vec{\gamma}\right)$, $\tilde{z}_{2} =\frac{1}{4} \left(\tilde{\sigma }\Lambda \left(\sigma +\Lambda \tilde{\sigma }\Lambda \right)^{-1} +\left(\sigma +\Lambda \tilde{\sigma }\Lambda \right)^{-1} \Lambda \tilde{\sigma }\right)\left(\vec{\varpi }+\Lambda \vec{\tilde{\varpi }}-\Lambda \tilde{\sigma }\vec{\gamma}\right)+\tilde{\sigma }\vec{\gamma}$.\\

The expressions for the Franck-Condon factors obtained above contain the modulus squared of multidimensional Hermite polynomials that can be computed using the recurrent relation derived in \cite{32}:
\begin{eqnarray} \label{eq4_7}
\nonumber\left|H_{\vec{n}\vec{m}} ^{\left\{R\right\}} \left(\vec{y}\right)\right|^{2} \\
=\left|\left(\sum _{j=1}^{N}r_{kj} y_{j}  \right)H_{\vec{v}-\vec{e}_{k} }^{^{\left\{R\right\}} } \left(\vec{y}\right)-\sum _{j\ne k}^{}r_{kj} v_{j}  H_{\vec{v}-\vec{e}_{k} -\vec{e}_{j} }^{^{\left\{R\right\}} } \left(\vec{y}\right)-r_{kk} \left(v_{k} -1\right)H_{\vec{v}-2\vec{e}_{k} }^{^{\left\{R\right\}} } \left(\vec{y}\right)\right|^{2},
\end{eqnarray}
where $\vec{v}=\left(n_1,...,n_N,m_1,..., m_N\right)$, $r_{kj}$ are the elements of the matrix $R$ and $\vec{e_k}$ is the $k$-th row of the identity matrix.

\section {Conclusion}

In the present work a new method of Franck-Condon factors calculation is proposed. It is based on the tomographic probability approach to quantum mechanics. Its advantage is that it requires to integrate fully positive and real function that is important for calculating multidimensional integrals having the sense of Franck-Condon factors in polyatomic molecules.\\

In terms of the tomographic method were considered particular cases corresponding to the change of an equilibrium position and a frequency of nuclei in diatomic and polyatomic molecules. The explicit relations for Franck-Condon factors in an arbitrary polyatomic molecule were obtained both in terms of the tomographic and the wave-function approaches.\\

\end{document}